\documentclass[aps,prb,reprint,twocolumn,superscriptaddress,amsmath,amssymb,letterpaper,floatfix,showpacs]{revtex4-1}

\usepackage{graphicx}% Include figure files

\begin{document}

\preprint{}

\title{Suppression of time reversal symmetry breaking superconductivity in Pr(Os,Ru)$_{\mathbf{4}}$Sb$_{\mathbf{12}}$ and (Pr,La)Os$_{\mathbf{4}}$Sb$_{\mathbf{12}}$}

\author{Lei Shu}
\altaffiliation{Present address: Department of Physics, University of California, San Diego, La Jolla, California 92093.}
\affiliation{Department of Physics and Astronomy, University of California, Riverside, California 92521}
\author{W. Higemoto}
\affiliation{Japan Atomic Energy Agency, Tokai-Mura, Ibaraki 319-1195, Japan}
\author{Y. Aoki}
\affiliation{Department of Physics, Tokyo Metropolitan University, Tokyo 192-0397, Japan}
\author{A.~D. Hillier}
\affiliation{ISIS facility, STFC Rutherford Appleton Laboratory, Harwell Science and Innovation Campus, Oxfordshire, OX11 0QX, United Kingdom}
\author{K. Ohishi}
\altaffiliation{Present address: Advanced Meson Science Laboratory, Nishina Center for Accelerator-Based Science, RIKEN, Wako 351-0198, Japan.}
\affiliation{Japan Atomic Energy Agency, Tokai-Mura, Ibaraki 319-1195, Japan}
\author{K. Ishida}
\affiliation{Department of Physics, Graduate School of Science, Kyoto University, Kyoto 606-8502, Japan}
\author{R. Kadono}
\author{A. Koda}
\affiliation{Meson Science Laboratory, KEK, Tsukuba, Ibaraki 305-0801, Japan}
\author{O. O. Bernal}
\affiliation{Department of Physics and Astronomy, California State University, Los Angeles, California 90032}
\author{D.~E. MacLaughlin}
\affiliation{Department of Physics, University of California, Riverside, California 92521}
\author{Y. Tunashima}
\author{Y. Yonezawa}
\author{S. Sanada}
\author{D. Kikuchi}
\author{H. Sato}
\affiliation{Department of Physics, Tokyo Metropolitan University, Tokyo 192-0397, Japan}
\author{H. Sugawara}
\altaffiliation{Present address: Department of Physics, Kobe University, Kobe 657-8501, Japan}
\affiliation{Faculty of the Integrated Arts and Sciences, The University of Tokushima, Tokushima 770-8502, Japan}
\author{T.U. Ito}
\affiliation{Japan Atomic Energy Agency, Tokai-Mura, Ibaraki 319-1195, Japan}
\author{M. B. Maple}
\affiliation{Department of Physics, University of California, San Diego, La Jolla, California 92093}

\date{\today}

\begin{abstract}
Zero-field muon spin relaxation experiments have been carried out in the Pr(Os$_{1-x}$Ru$_x$)$_4$Sb$_{12}$ and Pr$_{1-y}$La$_y$Os$_4$Sb$_{12}$ alloy systems to investigate broken time-reversal symmetry (TRS) in the superconducting state, signaled by the onset of a spontaneous static local magnetic field~$B_s$. In both alloy series $B_s$ initially decreases linearly with solute concentration. Ru doping is considerably more efficient than La doping, with a $\sim$50\% faster initial decrease. The data suggest that broken TRS is suppressed for Ru concentration $x \gtrsim 0.6$, but persists for essentially all La concentrations. Our data support a crystal-field excitonic Cooper pairing mechanism for TRS-breaking superconductivity.
\end{abstract}

% PACS, the Physics and Astronomy Classification Scheme.

\maketitle

In unconventional superconductors symmetries in addition to gauge symmetry are broken in the superconducting state, leading to novel properties and the possibility of more than one superconducting phase.~\cite{SiUe91} Breaking of time-reversal symmetry (TRS) by a superconducting transition is a relatively rare example of such additional broken symmetry. Strong experimental evidence for broken TRS comes from zero-field muon spin relaxation (ZF-$\mu$SR) experiments that observe the onset of a spontaneous local field~$B_s$ below the superconducting transition temperature~$T_c$. Spontaneous fields have been observed by ZF-$\mu$SR in (U,Th)Be$_{13}$,~\cite{HSWB90} UPt$_3$~\cite{LKLW93} (although not without controversy~\cite{DdRHYF95,HSNK00}), Sr$_2$RuO$_4$,~\cite{LFKL98} the first Pr-based heavy-fermion superconductor PrOs$_4$Sb$_{12}$,~\cite{ATKS03} LaNiC$_2$~\cite{HQC09} and, recently, PrPt$_4$Ge$_{12}$.~\cite{MSKG10} The ZF-$\mu$SR technique,~\cite{Brew94} in which spin-polarized muons are stopped in the sample and precess in their local fields, is very sensitive to small static fields and thus is ideally suited for the study of broken TRS in superconductors.

The isostructural filled-skutterudite compounds PrRu$_4$Sb$_{12}$  and LaOs$_4$Sb$_{12}$ are both conventional BCS-like superconductors ($T_{c} = 1.1$~K and 0.74~K, respectively).~\cite{TaIs00,BSFS01} Superconductivity is found for all values of Ru or La concentration in the alloy series Pr(Os$_{1-x}$Ru$_x$)$_4$Sb$_{12}$~\cite{FDHB04} and Pr$_{1-y}$La$_y$Os$_4$Sb$_{12}$,~\cite{RKA06} with relatively slow changes of $T_{c}$ with composition. This is quite different from the behavior of the majority of heavy-fermion superconductors, where chemical substitution rapidly suppresses $T_{c}$. In Pr(Os$_{1-x}$Ru$_x$)$_4$Sb$_{12}$ $T_{c}$ decreases smoothly from 1.85~K at $x = 0$ to a minimum of $\sim$0.75~K at $x \approx 0.6$, and then increases to 1.1~K at $x = 1$ (Ref.~\onlinecite{FDHB04}). In Pr$_{1-y}$La$_y$Os$_4$Sb$_{12}$ $T_{c}$ decreases monotonically with $y$ to 0.74~K at $y = 1$ (Ref.~\onlinecite{RKA06}). This behavior raises the question of how the TRS-breaking superconductivity of PrOs$_4$Sb$_{12}$ evolves with Ru and La substitution.

This Letter reports the results of ZF-$\mu$SR experiments in Pr(Os$_{1-x}$Ru$_{x}$)$_4$Sb$_{12}$ and Pr$_{1-y}$La$_{y}$Os$_4$Sb$_{12}$, which were undertaken to study the evolution of $\mathrm{B}_s$ with Ru and La doping. Preliminary results have been reported previously.~\cite{SHAF07} An initial linear decrease of $B_s$ with solute concentration is observed for both alloy series, but the data suggest very different effects of Ru and La: $B_s$ is suppressed $\sim$50\% faster by Ru doping than La doping and extrapolates to zero near the minimum in $T_c(x)$ (Ref.~\onlinecite{FDHB04}), whereas for La doping broken TRS appears to be present for most if not all La concentrations. Our results support the theory of TRS-breaking superconductivity from pairing via itinerant crystal-field excitations,~\cite{KMS06,Thal06} and motivate further studies of these systems.

The samples of PrOs$_4$Sb$_{12}$, Pr(Os$_{1-x}$Ru$_x$)$_4$Sb$_{12}$, and Pr$_{1-y}$La$_y$Os$_4$Sb$_{12}$ used in this study consist of randomly-oriented small ($\sim$0.1~mm) crystallites prepared by the Sb-flux method. Strong de Haas-van Alphen signals obtained from similarly-prepared crystals~\cite{SOSA02} attest to their high quality. ZF-$\mu$SR experiments were carried out at the Meson Science Laboratory, KEK, Tsukuba, Japan, and at the ISIS Neutron and Muon Facility, Rutherford Appleton Laboratory, Chilton, U.K.

\begin{figure}[ht]
\begin{center}
\includegraphics*[clip=,width=0.45\textwidth]{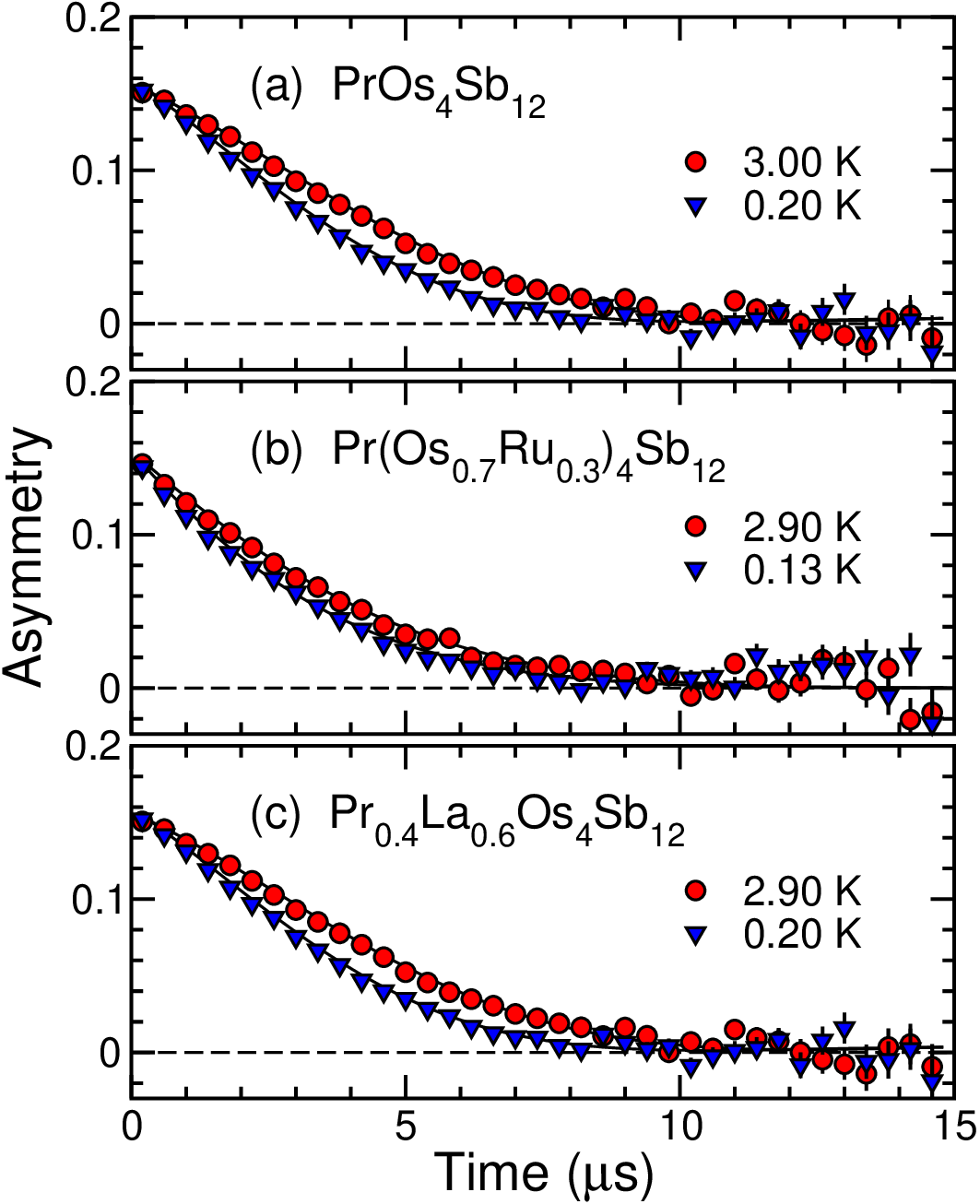}
\caption{(Color online) Time evolution of muon decay positron asymmetry, proportional to the muon spin polarization~$P_\mu(t)$, above and below the superconducting transition in PrOs$_4$Sb$_{12}$ and representative Ru- and La-based alloys.}
\label{fig:asy}
\end{center}
\end{figure}

Figure~\ref{fig:asy} shows the time evolution of the decay positron count rate asymmetry, proportional to the positive-muon ($\mu^+$) spin polarization~$P_\mu(t)$ (Ref.~\onlinecite{Brew94}), in PrOs$_4$Sb$_{12}$ and representative alloys at temperatures above and below $T_{c}$.
A constant background signal originating from muons stopping in the sample holder has been subtracted from the data. As previously reported,~\cite{ATKS03} in the end compound~PrOs$_4$Sb$_{12}$ the relaxation becomes faster in the superconducting state. Similar increases are observed in the alloys.

The ZF-$\mu$SR data are well described by the damped Gaussian Kubo-Toyabe (K-T) function~\cite{ATKS03,HUIN79}
\begin{equation}
 \label{eq:ds}
 P_{\mu}(t) = \exp(-\Lambda t)G_{z}^{\text{K-T}}(\Delta,t) ,
\end{equation}
where
\begin{equation}
 \label{eq:KT}
 G_{z}^{\text{K-T}}(\Delta,t)=\frac{1}{3}+\frac{2}{3}(1-\Delta^{2}t^{2})\exp(-{\textstyle\frac{1}{2}}\Delta^{2}t^{2})
\end{equation}
is the K-T functional form expected from an isotropic Gaussian distribution of randomly-oriented static (or quasistatic) local fields at muon sites.~\cite{HUIN79} The rms width of the static field distribution is $\Delta/\gamma_\mu$, where $\gamma_{\mu}$ is the muon gyromagnetic ratio, and $\Lambda$ is the exponential relaxation rate associated with an additional contribution to the muon spin relaxation. Both $\Delta$ and $\Lambda$ contribute to the increased low-temperature relaxation (Fig.~\ref{fig:asy}). 

\begin{figure}[ht]
\begin{center}
\includegraphics*[clip=,width=0.45\textwidth]{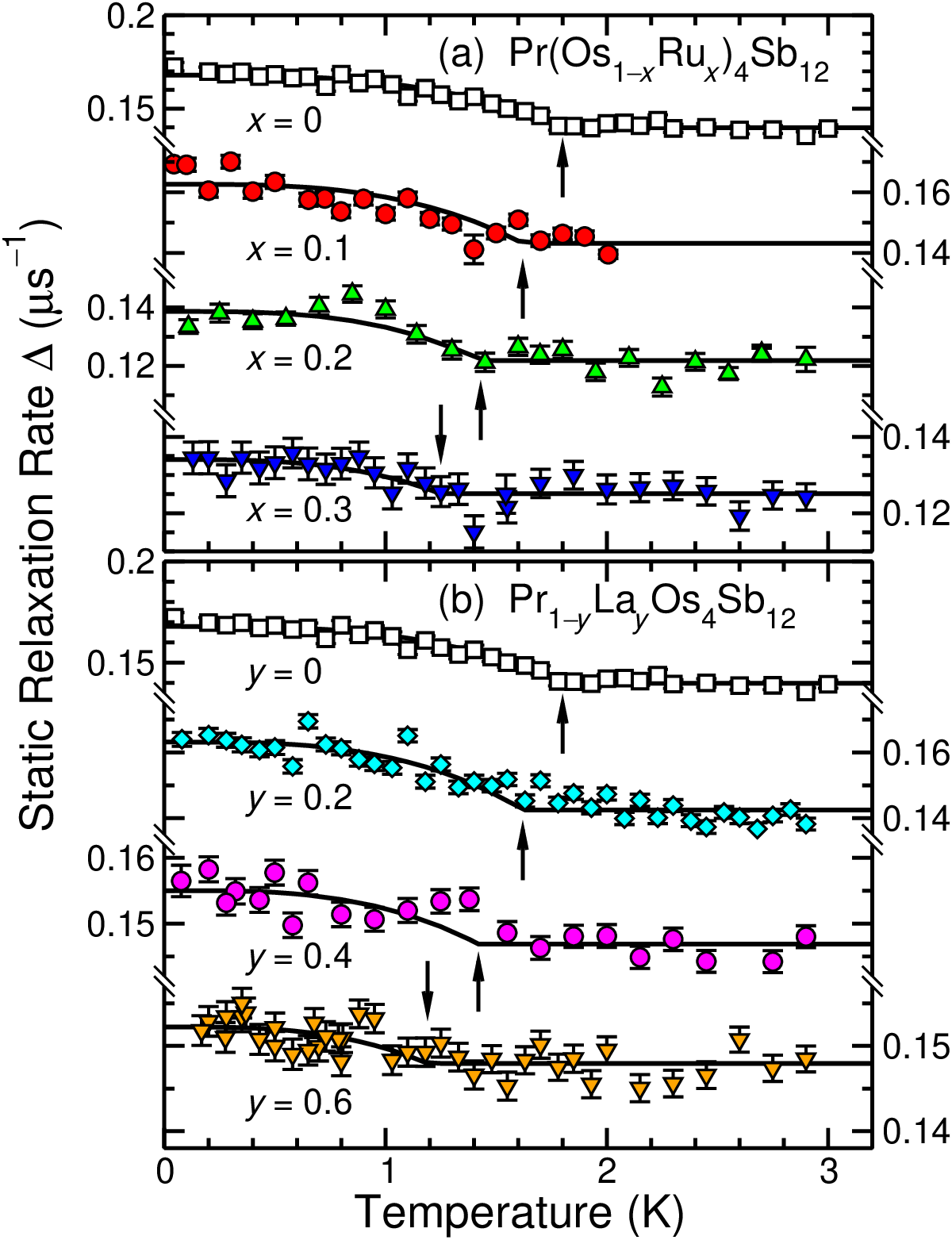}
\caption{(Color online) Points: temperature dependence of the ZF Kubo-Toyabe static relaxation rate~$\Delta$ in (a)~Pr(Os$_{1-x}$Ru$_x$)$_4$Sb$_{12}$ and (b)~Pr$_{1-y}$La$_y$Os$_4$Sb$_{12}$. Curves: fits of Eq.~(\protect\ref{eq:Delta}) to the data using the temperature dependence of the BCS order parameter for $\Delta_e(T)$ (barely visible for $x,y = 0$). Arrows: $T_{c}$ from bulk measurements.}
\label{fig:Del}
\end{center}
\end{figure}

Figure~\ref{fig:Del} shows the temperature dependence of $\Delta$ in Pr(Os$_{1-x}$Ru$_x$)$_4$Sb$_{12}$, $x = 0$, 0.1, 0.2, and 0.3, and Pr$_{1-y}$La$_y$Os$_4$Sb$_{12}$, $y = 0$, 0.2, 0.4, 0.6. 
In the normal state $\Delta$ is due to dipolar fields from neighboring nuclear magnetic moments. An increase in $\Delta$ below $T_{c}$ is observed in in both alloy series, indicating the onset of a spontaneous field in the superconducting state. The size of the increase becomes smaller with increasing solute concentration. To within errors no increase is observed in the end compounds PrRu$_4$Sb$_{12}$~\cite{AHPG05,SHAF07} and LaOs$_4$Sb$_{12}$.~\cite{AHSO05,SHAF07} In superconducting PrOs$_4$Sb$_{12}$~\cite{ATKS03} and Pr$_{0.8}$La$_{0.2}$Os$_4$Sb$_{12}$ longitudinal applied fields $\gtrsim H_{c1}$ (50--100~Oe) ``decouple'' the K-T relaxation, indicating that it is indeed quasistatic.~\cite{HUIN79} 

Below $T_{c}$ the nuclear dipolar and electronic contributions to $\Delta$ are uncorrelated and add in quadrature:~\cite{ATKS03}
\begin{equation}
 \label{eq:Delta}
\Delta(T) = [\Delta_{n}^{2}+\Delta_e^{2}(T)]^{1/2},
\end{equation}
where $\Delta_{n}$ is the normal-state nuclear dipolar rate and $\Delta_e(T)$ is the additional relaxation rate due to the spontaneous field from superconducting electrons. The K-T form assumes this field, like the nuclear dipolar field, is randomly distributed, although as previously noted~\cite{ATKS03} the data cannot discriminate between random and uniform spontaneous fields. A spontaneous internal field from broken TRS is expected only if (a)~the superconductor is inhomogeneous and the field is nonuniform,~\cite{SiUe91} or (b)~the pairing is nonunitary and the probe spin is hyperfine-coupled to the (uniform) Cooper-pair spin.~\cite{OhMa93} Thus it is not possible to decide between these alternatives from the ZF-$\mu$SR data alone.~\cite{ATKS03} 

\begin{figure}[ht]
\begin{center}
\includegraphics*[clip=,width=0.45\textwidth]{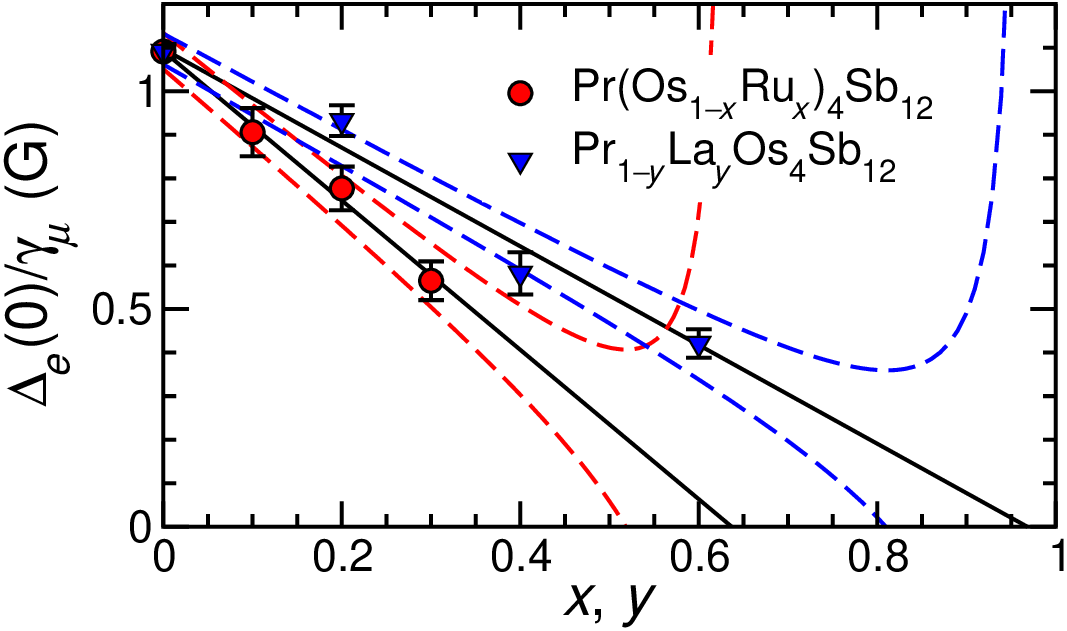}
\caption{(Color online) Dependence of the rms width~$\Delta_e(0)/\gamma_\mu$ of the $T = 0$ spontaneous field distribution on Ru concentration~$x$ and La concentration~$y$ in Pr(Os$_{1-x}$Ru$_x$)$_4$Sb$_{12}$ and Pr$_{1-y}$La$_y$Os$_4$Sb$_{12}$. Solid lines: linear fits. Dashed curves: estimated experimental error.}
\label{fig:deltae0}
\end{center}
\end{figure}

Equation~(\ref{eq:Delta}) was fit to the data of Fig.~\ref{fig:Del} using the temperature dependence of the BCS order parameter for $\Delta_e(T)$ (Ref.~\onlinecite{ATKS03}), and varying $\Delta_n$ and the amplitude $\Delta_e(0)$ of $\Delta_e(T)$ for best fit. Figure~\ref{fig:deltae0} shows the dependence of $\Delta_e(0)/\gamma_\mu$ on solute concentration.
The initial suppression is accurately linear for both solutes, with slopes~$-1.71(5)$~G (Ru doping) and $-1.13(7)$~G (La doping). For Ru doping the data are limited to $x \le 0.3$, and hence do not probe the crossover in penetration-depth behavior observed at higher concentrations.~\cite{CVSK05} The drastic decrease in the specific heat jump at $T_c$ observed in Pr$_{1-y}$La$_y$Os$_4$Sb$_{12}$ for $y \gtrsim 0.3$ (Ref.~\onlinecite{RKA06}) is not reflected in our data (Fig.~\ref{fig:deltae0}) [nor, for that matter, in $T_c(y)$].

The estimated error in $\Delta_e(0)$ diverges as $\Delta_e(0) \rightarrow 0$ (dashed curves in Fig.~\ref{fig:deltae0}). Thus data with experimentally attainable statistics (which are excellent, cf.\ Fig.~\ref{fig:asy}) cannot determine whether or not linearity is maintained at higher concentrations. Nevertheless the available data strongly suggest that broken TRS is suppressed for $x \gtrsim 0.6$ by Ru doping, but persists to high La concentrations. 

This behavior can be understood if the Cooper pairing mechanism in PrOs$_4$Sb$_{12}$ is the exchange of itinerant Pr$^{3+}$ crystal-field excitations (excitons).~\cite{MKH03,TKY04,Thal06,CETF07,Thal10} Treatments of this interaction have concluded that it can lead to TRS-breaking superconductivity in PrOs$_4$Sb$_{12}$~\cite{Thal06} and its alloys.~\cite{KMS06} An alternative picture for Pr(Os$_{1-x}$Ru$_x$)$_4$Sb$_{12}$ ~\cite{Thal10} explains the minimum in $T_c$ as due to a crossing between the CEF splitting and the rattling energy of the Pr ion~\cite{MAKS10} without considering broken TRS. 

In the model of Koga, Matsumoto, and Shiba (KMS),~\cite{KMS06} Ru and La doping of PrOs$_4$Sb$_{12}$ affect the TRS-breaking excitonic pairing in different ways. The rapid increase with Ru doping of the splitting between the crystalline-electric-field (CEF) Pr$^{3+}$ singlet ground state and magnetic first excited state~\cite{AYO06} weakens the pairing interaction due to the excitonic mechanism, without pair breaking or other effects that would rapidly suppress $T_c$. PrRu$_4$Sb$_{12}$ is a conventional superconductor, and this weakening leads to a crossover or transition between TRS-breaking superconductivity in Os-rich alloys and conventional $s$-wave pairing at the Ru-rich end of the series. This is reflected in the minimum in $T_c(x)$~\cite{FDHB04} and, as reported here, in the vanishing of the broken TRS, both for $x$ in the neighborhood of 0.6. 

In Pr$_{1-y}$La$_y$Os$_4$Sb$_{12}$ La substitutes for Pr with little distortion of the lattice or the electronic structure.~\cite{SOSA02,HaTa03,RKA06} La doping simply weakens the Pr-Pr intersite interaction, resulting in less excitonic dispersion; this reduces the pairing interaction.~\cite{KMS06} Our ZF-$\mu$SR data suggest that in this case TRS is broken  across the alloy series, with an amplitude that vanishes only for large $y$.

\begin{figure}[ht]
\begin{center}
\includegraphics*[clip=,width=0.45\textwidth]{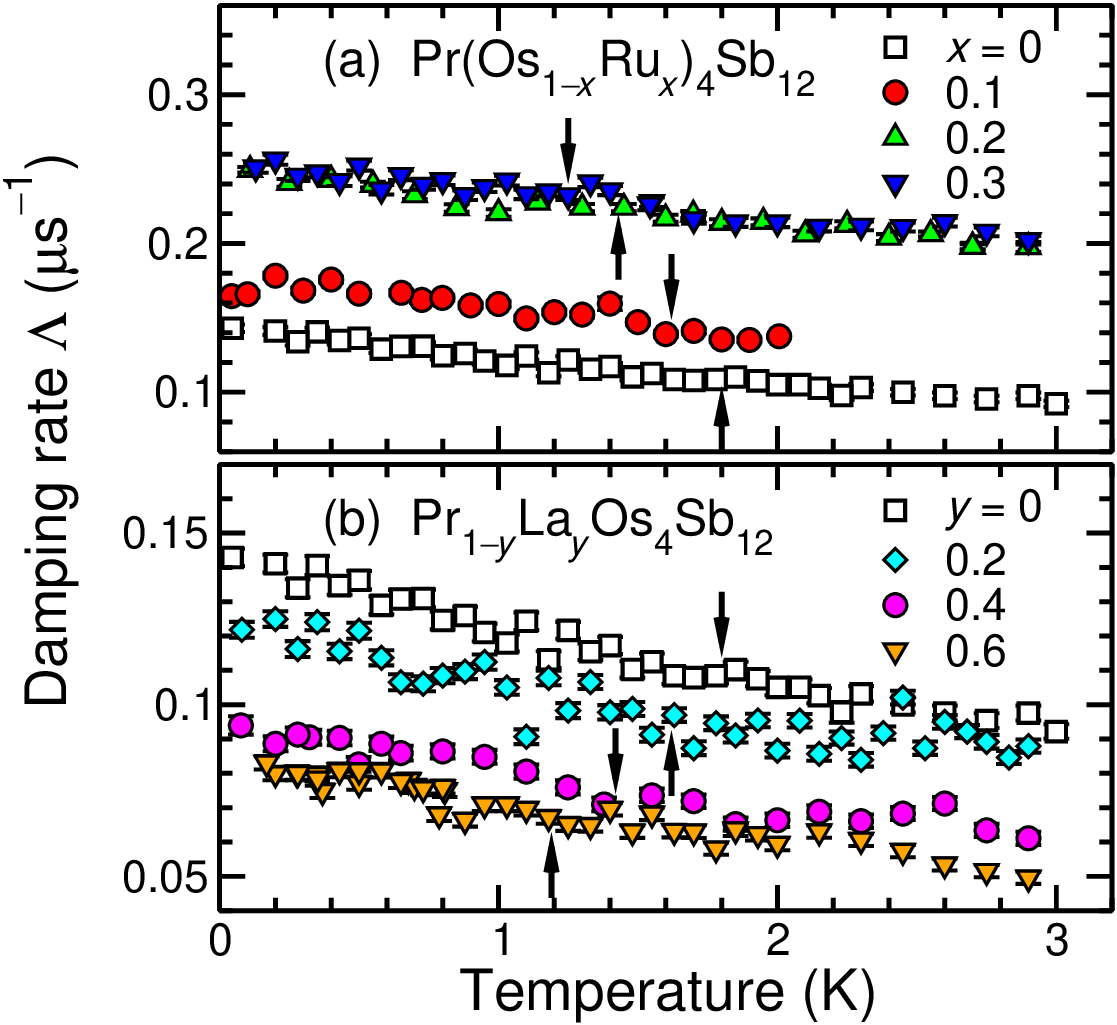}
\caption{(Color online) Temperature dependence of the ZF exponential damping rate~$\Lambda$ in (a)~Pr(Os$_{1-x}$Ru$_x$)$_4$Sb$_{12}$, $x = 0$, 0.1, 0.2, and 0.3, and (b)~Pr$_{1-y}$La$_y$Os$_4$Sb$_{12}$, $y = 0$, 0.2, 0.4, 0.6. Arrows: $T_{c}$ from bulk measurements.}
\label{fig:Lam}
\end{center}
\end{figure}

The temperature dependence of the exponential damping rate~$\Lambda$ is given in Fig.~\ref{fig:Lam} for the Ru-doped and La-doped alloys. 
As in PrOs$_4$Sb$_{12}$,~\cite{ATKS03} in the alloys there is no evidence for an anomaly in $\Lambda$ at $T_{c}$. Previous experiments~\cite{SMAT07} showed that the dependence of the damping on longitudinal field is consistent with dynamic relaxation due to thermal fluctuations. Nuclear magnetism was suggested as the origin of these fluctuations for a number of reasons, among them the fact that electronic spin fluctuations would be strongly affected by superconductivity.

We consider for completeness a transition to an unrelated weak-moment ($\lesssim 10^{-3}\mu_B$) magnetic state at $T = T_{\mathrm{mag}}$ as an alternative explanation of $B_s$. From the ZF-$\mu$SR data the onset of $B_s$ occurs at the superconducting $T_c$, at least for low doping where it can be clearly seen (Fig.~\ref{fig:Del}). Although there are many cases of coexistence of magnetism and superconductivity in strongly-correlated electron systems, $T_{\mathrm{mag}} = T_c$ only at isolated points in the phase diagrams of these systems. A magnetic-transition scenario requires fine tuning to such a point. Even if this were the case in the end compound, doping would almost certainly change $T_{\mathrm{mag}}$ relative to $T_c$, and the fine tuning would be lost in the alloys. There is no evidence for this, and we conclude that a magnetic transition unrelated to superconductivity is unlikely.

In Pr-based compounds the $\mu^+$ charge can affect the CEF splitting of Pr$^{3+}$ near neighbors, which in turn modifies the local Pr$^{3+}$ susceptibility that is the major contribution to the muon Knight shift.~\cite{[{See, for example, }][{ and references therein.}]TAGG97} This modification might also affect the superconductivity locally. The normal-state $\mu^{+}$ Knight shift in PrOs$_4$Sb$_{12}$ tracks the bulk susceptibility, however,~\cite{SMBH09} suggesting that any such perturbation is small.

We conclude that broken TRS in PrOs$_4$Sb$_{12}$ is suppressed by both Ru and La doping, but differently for the two solutes. Ru doping appears to restore TRS for $x \gtrsim 0.6$, near the minimum in $T_{c}(x)$, whereas for La doping TRS breaking persists to $y \sim 1$, i.e., most or all Pr-doped LaOs$_4$Sb$_{12}$ alloys exhibit broken TRS\@. These properties are consistent with the KMS picture~\cite{KMS06} for the CEF excitonic pairing mechanism and TRS-breaking superconductivity in PrOs$_4$Sb$_{12}$-based alloys. Our results motivate a quantitative treatment of broken TRS in Pr(Os$_{1-x}$Ru$_{x}$)$_4$Sb$_{12}$ and Pr$_{1-y}$La$_{y}$Os$_4$Sb$_{12}$.

We are grateful to the ISIS Cryogenics Group for invaluable help during the experiments, and to D.~F. Agterberg and C.~M. Varma for useful discussions. This work was supported by the U.S. NSF, grants 0422674 and 0801407 (Riverside), 0604015 (Los Angeles) and 0802478 (San Diego), the U.S. DOE, contract DE-FG-02-04ER46105 (San Diego), and by the Grant-in-Aid for Scientific Research on Priority Areas "Skutterudite" No. 15072206 and "Superclean Materials" No. 20029018, and on Innovative Areas "Heavy Electrons" No. 20102007 of the Ministry of Education, Culture, Sports, Science and Technology, Japan (Tokyo). 

%\bibliography{abbrev,comments,muSR,praseo,skutt,supercon,SrRuO,UThBe13}

%Merlin.mbs v4.21 2009-07-09.
%

\end{document}